\documentclass{tlp}

\usepackage{times}
\usepackage{soul}
\usepackage{url}
\usepackage[utf8]{inputenc}
\usepackage[small]{caption}
\usepackage{graphicx}
\usepackage{amsmath}
\usepackage{amssymb}
\usepackage{algorithm}
\usepackage{algorithmic}
\usepackage{booktabs}
\usepackage{multirow}
\usepackage[para]{threeparttable}

\usepackage[switch]{lineno} % Without this package, pdf was not compiling!
\usepackage{placeins} % to avoid Table 2 to appear among references
\usepackage{hyperref}

\newtheorem{defn}{Definition}

\newtheorem{cor}{Corollary}
\newtheorem{prop}{Proposition}

\def\ba{\begin{array}}
\def\ea{\end{array}}
\def\beq{\begin{equation}}
\def\eeq#1{\label{#1}\end{equation}}
\def\lar{\leftarrow}
\def\no{\ii{not}}
\def\ii#1{\hbox{\it #1\/}}
\def\clingo{{\sc Clingo}}
\def\choco{{\sc Choco}}

\usepackage{todonotes}

\begin{document}

\lefttitle{Yılmaz, Erdem}
\jnlPage{1}{8}
\jnlDoiYr{2021}
\doival{10.1017/xxxxx}

\title[Generating Satisfiable Benchmark Instances for SRTI with Optimization]{Generating Satisfiable Benchmark Instances for \\Stable Roommates Problems with Optimization}

\begin{authgrp}
\author{\sn{Baturay} \gn{Yilmaz}}
\affiliation{Faculty of Engineering and Natural Sciences, Sabanci University, Istanbul, Turkiye}
\author{\sn{Esra} \gn{Erdem}}
\affiliation{Faculty of Engineering and Natural Sciences, Sabanci University, Istanbul, Turkiye}
\end{authgrp}

\history{\sub{xx xx xxxx;} \rev{xx xx xxxx;} \acc{xx xx xxxx}}

\maketitle

\begin{abstract}
    While the existence of a stable matching for the stable roommates problem possibly with incomplete preference lists (SRI) can be decided in polynomial time, SRI problems with some fairness criteria are intractable. Egalitarian SRI that tries to maximize the total satisfaction of agents if a stable matching exists, is such a hard variant of SRI. For experimental evaluations of methods to solve these hard variants of SRI, several well-known algorithms have been used to randomly generate benchmark instances. However, these benchmark instances are not always satisfiable, and usually have a small number of stable matchings if one exists. For such SRI instances, despite the NP-hardness of Egalitarian SRI, it is practical to find an egalitarian stable matching by enumerating all stable matchings. In this study, we introduce a novel algorithm to generate benchmark instances for SRI that have very large numbers of solutions, and for which it is hard to find an egalitarian stable matching by enumerating all stable matchings. 
\end{abstract}

%%%%%%%%%%%%%%%%%%%%%%%%%%%%%%%%%%%%%%%%%%%%%%%%%%%%%%%%%%%%%%%%%%%%%%%%%%%%%

\section{Introduction}

Benchmark instances play an important role in empirical analysis of methods, and thus the generation of meaningful and/or hard benchmark instances have been studied for various problems in  AI~\citep{SelmanML96,Prosser14,TorralbaSS21,Dang2022} and OR~\citep{Pisinger05,VanhouckeM09}.  In this study, we are interested in generating hard and meaningful benchmark instances for the stable roommates problem possibly with incomplete preference lists (SRI)~\citep{GaleShapley62,Gusfield89}.

In SRI, each agent provides a strictly ordered and possibly incomplete list of their preferences over the other agents, and the goal is to find a roommate for each agent (if possible) so that no two agents prefer each other more than their current roommates. SRI (unlike the stable marriage problem) might admit no solutions~\citep{GaleShapley62}. The existence of a stable matching for SRI can be decided in polynomial time~\citep{Gusfield89}. On the other hand, SRI with some fairness criteria become intractable. For instance, Egalitarian SRI, which tries to maximize the total satisfaction of agents if a stable matching exists, is NP-hard~\citep{feder1992,Cseh19}.

For experimental evaluations of methods to solve hard optimization variants of SRI (like the ASP-based methods in our earlier study~\citep{Fidan20}), some well-known algorithms and tools have been used to randomly generate benchmark instances~\citep{GentP02,Prosser14}. For instance, we can randomly generate SRI instances using Prosser’s software~\citep{sricp2019} that is based on Mertens' idea~(\citeyear{Mertens2005}) to generate a random graph ensemble $G(n,p)$ according to the Erdös-Renyi model~(\citeyear{Erdos60}), where $n$ denotes the number of agents and $p$ denotes the probability of the mutual acceptability of pairs of agents. Although such benchmark generators are useful for understanding various aspects of methods to solve SRI instances, they may not be sufficient for a thorough and controlled experimental evaluation of methods for hard variants of SRI, due to the following challenges we have observed in our studies and real-world applications~\citep{fidan2021}.

\smallskip\noindent {\bf Challenge 1.} The SRI benchmark instances randomly generated by the existing methods usually have a small number of stable matchings if one exists. For instance, the benchmark instances generated for empirical analysis by \citeauthor{Fidan20}~(\citeyear{Fidan20}) have less than 4 stable matchings in average.  In such cases, an egalitarian stable matching can be found naively by enumerating all stable matchings. As Prosser notes~(\citeyear{Prosser14}):
{\it
``Therefore, although NP-hard, we would
fail to encounter a hard instance in the problems sampled. So, (as Cheeseman,
Kanefsky and Taylor famously asked~(\citeyear{Cheeseman91})) where are the hard problems?''}
With the existing methods, generating hard instances for optimization variants of SRI, in the spirit of Prosser's discussions (i.e., with too many stable matchings that is hard to enumerate), is challenging.

\smallskip\noindent {\bf Challenge 2.} SRI (unlike the stable marriage problem) might admit no solutions~\citep{GaleShapley62}, and the SRI benchmark instances randomly generated by the existing methods are not always satisfiable. For instance, almost half of the the benchmark instances generated for empirical analysis by \citeauthor{Fidan20}~(\citeyear{Fidan20}) for 80 agents, do not have a stable matching. On the other hand, for controlled empirical analysis, it is desirable to be able to generate instances that have stable matchings.  Ensuring the satisfiability of instances while generating them using the existing methods is challenging.

\smallskip\noindent {\bf Challenge 3.} To investigate the applicability of the proposed methods for SRI in the  real-world, it is desirable that the instances are also meaningful. For instance, in a real-world application of SRI at a university, it would not be meaningful to request long and strictly-ordered preference lists from students; usually, the preference lists would be short (e.g., containing 3--5 choices) and might include ties (SRTI). Generation of such meaningful SRTI instances by the existing methods is challenging.

\smallskip
In this study, motivated by these three challenges and the need for useful benchmark instances for empirical studies of SRTI, our contributions can be summarized as follows:

\smallskip\noindent {\bf Contribution 1.}
We introduce a method to generate hard benchmark instances for SRI possibly with Ties, that have large number of (e.g., over $10^6$) stable matchings, and for which it would not be practical to find an optimal (e.g., egalitarian) stable matching  by enumerating all stable matchings.

\smallskip\noindent {\bf Contribution 2.}
We introduce a ``seed \& combine" methodology underlying our method: (1) we generate small yet hard ``seed" instances with respect to desired features, and (2) we construct large instances by ``combining" these difficult seeds intelligently.

\smallskip\noindent {\bf Contribution 3.}
For (1), we utilize answer set programming (ASP)~\citep{MarekT99,Niemelae99,Lifschitz02,BrewkaET11,BrewkaEL16}---a declarative programming paradigm based on answer set semantics~\citep{gelfondL91,GelfondL88}, due to its expressive knowledge representation languages and efficient solvers. For~(2), we introduce a novel algorithm.
\smallskip

We illustrate the usefulness of our method with examples, and conclude with a discussion on its applicability to egalitarian one-to-one stable matching problems in general.

%%%%%%%%%%%%%%%%%%%%%%%%%%%%%%%%%%%%%%%%%%%%%%%%%%%%%%%%%%%%%%%%%%%%%%%%%%%%%

\section{Generating Solvable Hard Seed Instances}\label{sec:GeneratingSeed}

Our first objective is to generate ``seed" SRTI instances that are small (i.e., with small number of agents and with short preference lists) yet that are hard in the spirit of Prosser (i.e., that have very large number of stable matchings).

\subsection{Problem definition: {\sc seed-srti-sat}}

We consider the definition of SRTI and Egalitarian SRTI as in our earlier study~\citep{Fidan20}. Let us briefly remind the relevant definitions. 

An SRTI instance is characterized by a nonempty set $A$ of agents, and a collection~$\prec$ of preference lists $\prec_x$ of every agent $x$ in $A$. The preferences of an agent $x$ can be characterized by their ranks: for every agent $y$, if $y$ is the $i$'th preferred agent by $x$ then its rank $r(x,y)$ is $i$; here $1\leq r(x,y) \leq n-1$.  Therefore, if an agent $x$ prefers agent $y$ to agent $z$ then $r(x,y) < r(x,z)$. If an agent $x$ has the same preference for different agents $y$ and $z$ (i.e., $x$ is {\em indifferent} to $y$ and $z$) then $r(x,y){=}r(x,z)$ and the preference list $\prec_x$ contains a {\em tie} between $y$ and $z$. Note that if there is no tie in any preference list of any agent, then the problem becomes SRI.  

For every pair $\{x,y\}$ of different agents, if $x$ and $y$ prefer each other (i.e., $x\in\prec_y$ and $y\in \prec_x$), then $x$ and $y$ are {\em mutually acceptable} to each other.  Then a matching $\mu$ is characterized by the union of a set of pairs of mutually acceptable agents that are matched to each other, and a set of single agents. For every pair $\{x,y\}$ of agents that are mutually acceptable but not matched to each other in $\mu$, we say that $\{x,y\}$ {\em blocks} the matching $\mu$ if one of the following conditions hold: both agents are single in $\mu$, one of the agents $x$ is single and the other agent $y$ prefers $x$ to its current matched partner, or both agents are matched in $\mu$ but prefer each other to their current partners.  A matching for a given SRTI instance is {\em stable} if there is no such blocking pair of agents.  An SRTI instance that has a stable matching is {\em satisfiable}; otherwise, it is called {\em unsatisfiable}.

Now, let us proceed with our definitions. We define a satisfiable seed SRTI instance as follows.

\begin{defn}[Satisfiable seed SRTI instances]\label{def:SatisfiableSeedInstance}
Given positive integers $n{>}1$, $m{<}n$ and $k$, a {\em satisfiable $(n,m,k)$-seed SRTI instance} $(I,S_I)$ is a pair of 
 \begin{itemize}
  \item 
  an SRTI instance $I{=}(A,\prec)$ for a nonempty set $A$ of agents with preference lists $\prec$ where $|A|{=}n$, and $|\prec_x|{\leq} m$ for every agent $x$ in $A$, and
  \item 
  a set $S_I$ of exactly $k$ different stable matchings for $I$.
 \end{itemize}
\end{defn}

We are interested in the problem of generating such satisfiable seed SRTI instances to construct benchmarks for SRTI.

\begin{defn}[{\sc seed-srti-sat}]\label{def:dec-seed}
Given positive integers $n{>}1$, $m{<}n$ and $k$, {\sc seed-srti-sat} is the problem of deciding the existence of a satisfiable $(n,m,k)$-seed SRTI instance.
\end{defn}

We generate such satisfiable $(n,m,k)$-seed SRTI instances, using ASP.

\subsection{Solving {\sc seed-srti-sat} using ASP}
\label{sec:seed-asp}

Our ASP formulation of {\sc seed-srti-sat} consists of three parts. The first part generates preference lists of length at most $m$, for each agent. Given such preference lists, the second part generates $k$ different matchings and the third part ensures that each matching is stable.

\paragraph{\bf Part 1: Generating preference lists}
Suppose that the input is described by positive integers $n{>}1$, $m{<}n$ and~$k$.

\underline{First}, for each agent, we generate a preference list of length at most $m$. Suppose that a preference list $\prec_{a_1}$ of an agent $a_1$ is described by atoms of the form $\ii{arank}(a_1,a_2,r)$ (``agent $a_2$ has rank $r$ in the preference list $\prec_{a_1}$ of $a_1$'').

For each agent $a_1$ and for each rank $r$,
we nondeterministically choose a subset of agents $a_2$
and include them in the preference list of $a_1$ at rank $r$, by the choice rule:
\[
\ba l
\{\ii{arank}(a_1,a_2,r): 1{\leq} \ a_2 {\leq} n,\ a_1 {\neq} a_2\}. %\\
\quad (1{\leq} \ a_1 {\leq} n;\ 1{\leq} r {\leq} m)
\ea
\]
We ensure that $a_2$ has a unique rank $r$ in the preference list of $a_1$ by the constraints:
\[
\ba l
\lar \ii{arank}(a_1,a_2,r), \ii{arank}(a_1,a_2,r_1). %\\
\quad (1{\leq} \ a_1,a_2 {\leq} n; a_1 {\neq} a_2; 1{\leq} r, r_1 {\leq} m; r {\neq} r_1)
\ea
\]
Here we also make sure that the ranks start from 1 and are consecutive:
\[
\ba l
\lar \no\ \ii{arank}(a,\_,r-1), \ii{arank}(a,\_,r). %\\
\quad (1{\leq} \ a {\leq} n; 2 {\leq} r {\leq} m)
\ea
\]
Lastly, we ensure that the preference lists contains at most $m$ agents:
\[
\ba l
\lar \texttt{\#}count\{a_2 : arank(a_1,a_2,\_)\} {>} m. %\\ 
\quad (1{\leq} a_1 {\leq} n)
\ea
\]

\paragraph{\bf Part 2: Generating $k$ different matchings}
We ensure that the SRTI instance $(A,\prec)$ generated above has $k$ matchings.

\underline{First}, we define the mutual acceptability of two agents $a_1$ and $a_2$, and the preference of an agent $a_1$ over another agent $a_2$ by an agent $a$ by the rules:
\[
\ba l
\ii{acceptable}(a_1,a_2) \lar \ii{arank}(a_1,a_2,\_), \ii{arank}(a_2,a_1,\_). %\\
\quad (1{\leq} \ a_1,a_2 {\leq} n; a_1 {\neq} a_2) \\
\ii{aPrefers}(a,a_1,a_2) \lar \ii{arank}(a,a_1,r_1), \ii{arank}(a,a_2,r_2). \\
\quad (1{\leq} \ a,a_1,a_2 {\leq} n; a_1 {\neq} a_2;1{\leq} r_1, r_2 {\leq} m; r_1{<}r_2) \\
\ea
\]

\underline{Next}, we generate $k$ matchings. 
Suppose that atoms of the form $\ii{matched}(a_1,a_2,i)$ denote that agent $a_1$ is assigned to agent $a_2$ in matching~$i$.
At every matching $i$, for every acceptable pair of agents $a_1$ and $a_2$, we nondeterministically decide to assign $a_2$ to $a_1$:
\[
\ba l
\{\ii{matched}(a_1,a_2,i) : \ii{acceptable}(a_1,a_2)\}1.  %\\
\quad (1{\leq} a_1,a_2 {\leq} n;\ a_1 {\neq} a_2;\ 1{\leq} i {\leq} k)
\ea
\]
We ensure that this assignment is symmetric:
\[
\ba l
\lar \ii{matched}(a_1,a_2,i), \no\ \ii{matched}(a_2,a_1,i). %\\
\quad (1{\leq} a_1,a_2 {\leq} n;\ 1{\leq} i {\leq} k)
\ea
\]
and that an agent $a$ cannot be assigned to two different agents $a_1$ and $a_2$:
\[
\ba l
\lar \ii{matched}(a,a_1,i), \; \ii{matched}(a,a_2,i). 
\quad (1{\leq} a,a_1,a_2 {\leq} n;\ a {\neq} a_1,a_2;\ a_1 {\neq} a_2;\ 1{\leq} i {\leq} k)
\ea
\]

\underline{Next}, for every matching $i$, we identify every agent $a$ who is not assigned to another agent:
\[
\ba l
\ii{aSingle}(a,i) \lar \no\ \ii{matched}(a,\_,i). %\\
\quad (1{\leq} a{\leq} n;\ 1{\leq} i{\leq} k)
\ea
\]

\underline{Next}, we ensure that these $k$ matchings are different from each other.
For two matchings $i_1$ and $i_2$, first we define when they are different: if there exists an agent $a$ who is assigned to different agents $a_1$ and $a_2$ in them.
\[
\ba l
\ii{hasDifferentPairs}(i_1,i_2) \lar %\\ \qquad 
\ii{matched}(a,a_1,i_1), \ii{matched}(a,a_2,i_2). \\
\quad (1{\leq} a,a_1,a_2 {\leq} n;\ a {\neq} a_1,a_2;\ a_1 {\neq} a_2;\ 1{\leq} i_1,i_2 {\leq} k; i_1{\neq} i_2) \\
\ii{hasDifferentPairs}(i_1,i_2) \lar %\\\qquad 
\ii{matched}(a,\_,i_1), \ii{aSingle}(a,i_2). \\
\quad (1{\leq} a {\leq} n;\ 1{\leq} i_1,i_2 {\leq} k; i_1{\neq} i_2)
\ea
\]
\underline{Then}, we ensure that any two of the $k$ matchings generated above are different:
\[
\ba l
\lar \no\ \ii{hasDifferentPairs}(i_1,i_2). %\\
\quad (1{\leq} i_1,i_2 {\leq} k; i_1{\neq} i_2)
\ea
\]

\paragraph{\bf Part 3: Ensuring stability}

We need to guarantee that every generated unique matching $i$ above is stable.

\underline{First}, for every matching $i$, we identify every pair $(a_1,a_2)$ of agents that blocks the matching~$i$:
\[\ba l
\ii{blockingPair}(a_1,a_2,i) \lar 
\ii{aSingle}(a_1,i), \ii{aSingle}(a_2,i), \\ \qquad
\ii{acceptable}(a_1,a_2). 
\quad (1{\leq} a_1,a_2 {\leq} n;\ a_1 {\neq} a_2;\ 1{\leq} i{\leq} k) \\
\ii{blockingPair}(a_1,a_2,i) \lar 
\ii{aSingle}(a_2,i), \ii{matched}(a_1,x,i), 
\ii{aPrefers}(a_1,a_2,x), \\ \qquad 
\ii{acceptable}(a_1,a_2).
\quad (1{\leq} x,a_1,a_2 {\leq} n;\ x {\neq} a_1,a_2;\ a_1 {\neq} a_2;\ 1{\leq} i{\leq} k) \\
\ii{blockingPair}(a_1,a_2,i) \lar 
\ii{aSingle}(a_1,i), \ii{matched}(a_2,x,i), 
\ii{aPrefers}(a_2,a_1,x), \\ \qquad 
\ii{acceptable}(a_1,a_2).
\quad (1{\leq} x,a_1,a_2 {\leq} n;\ x {\neq} a_1,a_2;\ a_1 {\neq} a_2;\ 1{\leq} i{\leq} k) \\
\ii{blockingPair}(a_1,a_2,x) \lar 
\ii{matched}(a_1,x,i), \ii{matched}(a_2,y,i), 
\ii{aPrefers}(a_1,a_2,x), \\ \qquad 
\ii{aPrefers}(a_2,a_1,y). 
\quad (1{\leq} x,y,a_1,a_2 {\leq} n;\ x,y {\neq} a_1,a_2;\ a_1 {\neq} a_2;\ 1{\leq} i{\leq} k) \\
\ea\]

\underline{Then}, we ensure that, for every matching $i$, there is no pair of agents $a_1$ and $a_2$ that blocks the matching~$i$:
\[
\ba l
\lar \ii{blockingPair}(a_1,a_2,i). %\\
\quad (1{\leq} a_1,a_2{\leq} n;\ a_1{\neq} a_2;\ 1{\leq} i{\leq} k)
\ea
\]

\noindent{\bf Example} Using the ASP program above, we can generate small yet hard SRTI instances.
For instance, a seed SRTI instance with $n{=}4$ agents, $m{=}n{-}1$ and $k{=}2$ stable matchings, generated by our program is illustrated in the upper left table (SRTI~1) in Figure~\ref{fig:example}.

\begin{figure*}
\centering
\resizebox{\textwidth}{!}{
\begin{oldtabular}{cc}
  \\  SRTI 1: \\
        \begin{oldtabular}{ll}
            \hline
            $a_1$:& ($a_4$)
            \\
            $a_2$:& ($a_4$)
            \\
            $a_3$:& ($a_2$)
            \\
            $a_4$:& ($a_3$) ($a_1$, $a_2$)
            \\\hline
        \end{oldtabular}
        &
        \begin{oldtabular}{ll}
            \hline
            $\mu_1^1$:& $\{(a_1, a_4), (a_2), (a_3)\}$
            \\
            $\mu_1^2$:& $\{(a_2, a_4), (a_1), (a_3)\}$
            \\\hline
        \end{oldtabular}
  \\ \\  \\ SRTI 2: \\
        \begin{oldtabular}{ll}
            \hline
            $a_5$:& ($a_9$)
            \\
            $a_6$:& ($a_8$) ($a_7$) ($a_9$)
            \\
            $a_7$:& ($a_9$)
            \\
            $a_8$:& ($a_7$)
            \\
            $a_9$:& ($a_5$, $a_6$, $a_7$)
            \\\hline
        \end{oldtabular}
        &
        \begin{oldtabular}{ll}
            \hline
            $\mu_2^1$:& $\{(a_7, a_9), (a_5), (a_6), (a_8)\}$
            \\
            $\mu_2^2$:& $\{(a_6, a_9), (a_5), (a_7), (a_8)\}$
            \\
            $\mu_2^3$:& $\{(a_5, a_9), (a_6), (a_7), (a_8)\}$
            \\\hline
        \end{oldtabular}
\\  \\  \\ SRTI 3: \\
        \begin{oldtabular}{ll}
            \hline
            $a_1$:& ($a_7$, $a_9$) ($a_4$)
            \\
            $a_2$:& ($a_4$)
            \\
            $a_3$:& ($a_2$) ($a_9$)
            \\
            $a_4$:& ($a_3$) ($a_1$, $a_2$) ($a_7$) ($a_9$) ($a_8$)
            \\
            $a_5$:& ($a_4$, $a_9$) ($a_3$)
            \\
            $a_6$:& ($a_8$) ($a_4$) ($a_7$) ($a_9$)
            \\
            $a_7$:& ($a_9$) ($a_4$)
            \\
            $a_8$:& ($a_2$, $a_3$, $a_7$) ($a_4$)
            \\
            $a_9$:& ($a_2$, $a_3$, $a_5$, $a_6$, $a_7$) ($a_4$)
            \\\hline
        \end{oldtabular}
        &
        \begin{oldtabular}{ll}
            \hline
            $\mu_1$:& $\{(a_1, a_4), (a_7, a_9), (a_2), (a_3), (a_5), (a_6), (a_8)\}$
            \\
            $\mu_2$:& $\{(a_1, a_4), (a_6, a_9), (a_2), (a_3), (a_5), (a_7), (a_8)\}$
            \\
            $\mu_3$:& $\{(a_1, a_4), (a_5, a_9), (a_2), (a_3), (a_6), (a_7), (a_8)\}$
            \\
            $\mu_4$:& $\{(a_2, a_4), (a_7, a_9), (a_1), (a_3), (a_5), (a_6), (a_8)\}$
            \\
            $\mu_5$:& $\{(a_2, a_4), (a_6, a_9), (a_1), (a_3), (a_5), (a_7), (a_8)\}$
            \\
            $\mu_6$:& $\{(a_2, a_4), (a_5, a_9), (a_1), (a_3), (a_6), (a_7), (a_8)\}$
            \\
            $\mu_7$:& $\{(a_2, a_4), (a_3, a_9), (a_1), (a_5), (a_6), (a_7), (a_8)\}$
            \\
            $\mu_8$:& $\{(a_1, a_4), (a_3, a_9), (a_2), (a_5), (a_6), (a_7), (a_8)\}$
            \\\hline
        \end{oldtabular}
\\ \\   \\
\end{oldtabular}}
      \caption{(upper left) First small seed SRTI instance, $I^1$, (upper right) 2 stable matchings of the first seed, $S_{I^1}$, (middle left) second small SRTI seed with renamed agents, $I^2$, (middle right) stable matchings of the second seed, $S_{I^2}$, (lower left) an SRTI instance obtained from two seeds, $\mathbf{I}$, (lower right) 8 stable matchings of the combined instance.}
    \label{fig:example}
\end{figure*}

%%%%%%%%%%%%%%%%%%%%%%%%%%%%%%%%%%%%%%%%%%%%%%%%%%%%%%%%%%%%%%%%%%%%%%%%%%%%%
\section{Populating and Combining Seeds into Large SRTI Instances}\label{sec:CombiningSeeds}

First we generate $c$ $(n_i,m_i,k_i)$-seed SRTI instances $(I^i,S_{I^i})$ ($1{\leq}i{\leq}c$), where $I^i=(A^i,\prec^i$), possibly with different values of $n_i$, $m_i$ and $k_i$. Then we rename the agents in such a way that no agent appears in two seed instances.

Next, we ``combine" these $c$ seeds into a larger SRTI instance $\mathbf{I}$, in such a way that some desired properties of the seeds (e.g., short preference lists) are preserved in $\mathbf{I}$, and that the instance $\mathbf{I}$ is hard (i.e., the number of stable matchings for $\mathbf{I}$ is at least $\prod_{i{=}1}^ck_i$). 

Essentially, for some pairs $(x,y)$ of agents such that $x$ and $y$ appear in different seeds $I^x$ and $I^y$, respectively, our method updates their preference lists by adding $x$ into the preference list of~$y$ with a probability $p_1$ of incompleteness and with a probability $p_2$ of forming a tie.  In this way, we ``combine'' the seeds $I^x$ and $I^y$, and form a new and larger SRTI instance.

A trial of adding~$x$ in the preference list of $y$ in $I^y$ with respect to $p_1$ and $p_2$ is described as follows:
    \begin{enumerate}
    \item
    First, observe that $x$ cannot be added to the preference list of~$y$ if the length of the preference list of $y$ is already~$m$, or if $x$ is already in the preference list of $y$.
    
    Otherwise, if the length of the preference list of $y$ is strictly less than~$m$ and if $x$ is not already in the preference list of $y$ in~$I^y$, randomly pick two numbers $r_1$ and $r_2$ in $[0,1]$ and proceed with the following steps. 
    
    \item   
    If $r_1 {>} p_1$ and $y$ is not in the preference list of $x$ in $I^x$, then add $x$ to the preference list of $y$ in $I^y$ as follows. 
    
    If $r_2 {<} p_2$, then add $x$ to the preference list of $y$ in $I^y$ so that $y$ becomes indifferent between $x$ and at least one other agent. 
    
    Otherwise, add $x$ to the preference list of $y$ in $I^y$ so that $x$ does not appear in a tie.
    
    \item If $r_1 {>} p_1$ and $y$ is already in the preference list of $x$ in $I^x$, first observe that adding $x$ to the preference list of $y$ in $I^y$ will create a mutually acceptable pair $(x,y)$ for the combined instance. 
    So we can try to add $x$ to the preference list of $y$ at some rank, provided that $(x,y)$ will not block any stable matching of the seed instances.  
    
    If $y$ is single in a matching $\mu_y^j \in S_{I^y}$ and, $x$ is single in a matching $\mu_x^i \in S_{I^x}$ or $x$ prefers $y$ to at least one his matched partner, then $(x,y)$ would block the matchings of the combined instance. 
    
    Otherwise, if $y$ is matched in every matching in $S_{I^y}$ or, $x$ is matched in every matching $S_{I^x}$ and $x$ does not prefer $y$ to none of his matched partners, then add $x$ to the preference list of $y$ in $I^y$ as described in Step~2 above, but with a rank that is greater than or equal to the ranks of every partner of $y$ in the matchings in $S_{I^y}$ (when $x$ is matched in every matching in $S_{I^x}$). 
    
    Observe that, if $x$ is matched in every matching in $S_{I^x}$ and $y$ is single in at least one matching, then we can put $x$ at any rank in the preference list of $y$. Because, since $x$ and $y$ are in two different seeds and $y$ is in the preference list of $x$, then $y$ must have been added to that preference list at some point by this method. Also, since $y$ is single at some matching, it must succeed every partner of $x$ in the preference list as required by the method. Moreover, observe that, with this rank, since one of $x$ and $y$ is the least preferred agent for the other, $(x,y)$ does not block any stable matching of the seed instances. 

    \item In the end, after considering every possible pair $(x,y)$ where $x$ and $y$ occur in different seeds, our method constructs an SRTI instance $\mathbf{I}=(A^x \cup A^y, \prec')$ where $\prec'$ is the collection that contains updated preferences lists. Note that if $x$ cannot be added to the preference list of $y$, then the combined instance is still generated containing both agents $x$ and $y$ without including $x$ in the preference list of $y$.
    \end{enumerate}

 Our method of combining two seed instances over a pair of agents probabilistically, preserves the stable matchings of the seed instances. 

\begin{prop} \label{prop:combine}
Given a $(n_x,m_x,k_x)$-seed instance $(I^x, S_{I^x})$ and 
a $(n_y,m_y,k_y)$-seed instance $(I^y, S_{I^y})$ with unique agents, two agents $x$ and $y$ of $I^x$ and $I^y$ respectively, and probabilities $p_1$ and $p_2$, let $\mathbf{I}$ be an SRTI instance obtained from these seeds by our seed \& combine method above.  For every stable matching $\mu_x \in S_{I^x}$ and for every stable matching $\mu_y \in S_{I^y}$, $\mu_x \cup \mu_y$ is a stable matching for $\mathbf{I}$.
\end{prop}

%\begin{proof}[\bf Proof of Proposition~\ref{prop:combine}]
\noindent{\bf Proof} According to our method, for two agents $x$ and $y$ that belong to two different seeds, there can be three cases. At the end of the procedure, either $y$ does not put $x$ into its preference list, or $y$ puts $x$ into its preference list, but $(x,y)$ is not mutually acceptable, or $y$ puts $x$ into its preference list and they become mutually acceptable.

    \begin{itemize}
        \item[] \textit{\underline{Case 1:} $x$ is not added to the preference list of $y$.} Observe that, since $\mu_x$ and $\mu_y$ are stable by definition, clearly, no pair in $I^x$ (respectively $I^y$) blocks $\mu_x$ (respectively $\mu_y$). Also, observe that no pair in $I^x$ (respectively $I^y$) can block $\mu_y$ (respectively $\mu_x$); otherwise, by definition, a member in the blocking pair in $I^x$ (respectively $I^y$) must be included in $I^y$ (respectively $I^x$). However, agents in the seeds are given to be unique, so it is not possible. So, $\mu_x \cup \mu_y$ is stable as it is. Also, since there is no new pair $(x,y)$ that is mutually acceptable, there can be no blocking pair. Then, $\mu_x \cup \mu_y$ remains stable.

        \item[] \textit{\underline{Case 2.1:} $x$ is added to the preference list of $y$ and $(x,y)$ is not mutually acceptable.} This case corresponds to Step~2 of the method above. In this case, $x$ is added to the preference list of $y$; however, since $(x,y)$ is not an acceptable pair, then it is also not blocking by definition. Thus, $\mu_x \cup \mu_y$ is stable.

        \item[] \textit{\underline{Case 2.2:} $x$ is added to the preference list of $y$ and $(x,y)$ is mutually acceptable.} This case corresponds to Step~3 of the method above. In this case, since $x$ is added to the preference list of $y$, we know that they are not simultaneously single at $\mu_x, \mu_y$, and at least one of them prefers all of its matched partners to the other one. Then, by definition, the pair $(x,y)$ cannot be a blocking pair, and hence $\mu_x \cup \mu_y$ is stable.
        
    \end{itemize}
    Since $\mu_x \cup \mu_y$ is stable for all of the cases, it must also be stable for $\mathbf{I}$.
% $\qed$
\hspace{1em} $\square$
%\end{proof}

This result generalizes to combining more than two seeds, by repeated application of our method.

\begin{cor} 
Given $c$ $(n_i,m_i,k_i)$-seed SRTI instances $(I^i, S_{I^i})$ ($1{\leq}i{\leq}c$) with unique agents, and probabilities $p_1$ and $p_2$, let $\mathbf{I}$ be an SRTI instance obtained by combining pairs of these seeds by using our method above. Let $\mathcal{M}(\mathbf{I})$ be the set of stable matchings of $\mathbf{I}$. Then, $S_{I^1} \times S_{I^2} \times \dots \times S_{I^c} \subseteq \mathcal{M}(\mathbf{I})$.
\end{cor} 

\paragraph{\bf Example (cont'd)}
Figure~\ref{fig:example} illustrates two small seed SRTI instances: SRTI 1 and SRTI 2. SRTI~1 is a (4,3,2)-seed SRTI instance specified by $I^1$ defined over the agents $A_1{=}\{a_1,a_2,a_3,a_4\}$ and their preferences $\prec^1$ (upper left), and by $S_{I^1}$ consisting of 2 stable matchings for $I^1$ (upper right: $\mu_1^1$ and $\mu_2^1$).

SRTI 2 is a (5,4,3)-seed SRTI instance specified by $I^2$ defined over the agents $A_2{=}\{a_5,a_6,a_7,a_8\}$ and their preferences $\prec^2$ (middle left), and by $S_{I^2}$ consisting of 3 stable matchings for $I^2$ (middle right: $\mu_1^2$,  $\mu_2^2$, $\mu_3^2$).

SRTI 3 is an SRTI instance obtained from these two different seed instances, considering $p_1= 0.5$ and $p_2=0.5$. In particular, our method tries to add each agent $a_1..a_4$ to the preference lists of each agent $a_5..a_9$, and vice versa. 

For example, $a_7$ is added in a tie in the preference list of $a_1$, while it is added to the preference list of $a_4$ without being in a tie. 
On the other hand, $a_1$ could not be added to the preference list of $a_7$, to prevent $(a_1,a_7)$ from blocking the stable matchings $\mu_5$ and $\mu_6$ of SRTI 3. 

Our method preserves the stable matchings of the seed instances:  Observe that every element of $S_{I^1} \times S_{I^2}$ (i.e., each one of the first 6 matchings in the lower right table) is a stable matching for SRTI 3. 

On the other hand, our method does not preserve the ranks of agents while combining the seeds.  For example, $a_4$ is in the first rank at the preference list of $a_1$ in SRTI 1, while it is in the second rank at SRTI 3. 

Also, note that an SRTI instance obtained from two seed instances, $(I^1,S_{I^1})$ and $(I^2,S_{I^2})$, can have more than $|S_{I^1}|*|S_{I^2}|$ stable matchings. For example, SRTI 3 has 2 more stable matchings, $\mu_7$ and $\mu_8$. This example also shows why the claim of Proposition~\ref{prop:combine} is not bidirectional. 

%%%%%%%%%%%%%%%%%%%%%%%%%%%%%%%%%%%%%%%%%%%%%%%%%%%%%%%%%%%%%%%%%%%%%%%%%%%
\section{Remarks}
 \paragraph{\bf Edge cases} Our method considers two probabilities given in the input: $p_1$ (probability of incompleteness) and $p_2$ (probability of ties).  We would like to underline that, in the edge cases where $p_1$ or $p_2$ is given as $0$ or $1$, the preference lists of satisfiable seed SRTI instances are constructed carefully. 
 
 For instance, if the user gives $p_1=0$ as input, then the user tries to generate an instance where there is no incompleteness (i.e., the preference lists must be complete, including all $n-1$ agents). Hence, the generated seed instance must also satisfy this property. We ensure this by adding the following constraint to the ASP program used for generating seed instances (Section~\ref{sec:seed-asp}).
\[
\lar  \texttt{\#}\ii{count}\{a_2 : \ii{arank}(a_1,a_2,\_)\} {\neq} n-1. \quad (1 {\leq} a_1 {\leq} n)
\]

If the user specifies $p_1=1$ in the input, then every preference list must be empty. Thus there is no need to generate an instance.  

If the user specifies $p_2=0$ in the input, then the user tries to generate an instance without any ties in the preference lists. We ensure this by adding the following constraint to our seed generating ASP program.
\[
\lar  \ii{arank}(a,a_1,r), \ii{arank}(a,a_2,r). \quad (1{\leq} a,a_1,a_2 {\leq} n;\ a_1 {\neq} a_2;\ 1{\leq} r {\leq} m)
\]

If the user specifies $p_2=1$ in the input, then every agent must prefer other agents at the same rank. In other words, it is not allowed that an agent prefers different agents at different ranks. We ensure this by adding the following constraint to our seed generating ASP program.
\[
\lar  \ii{arank}(a,a_1,r_1), \ii{arank}(a,a_2,r_2). \quad (1{\leq} a,a_1,a_2 {\leq} n;\ a_1 {\neq} a_2;\ 1{\leq} r_1,r_2 {\leq} m;\ r_1 {\neq} r_2)
\]

Our seed \& combine algorithm (Section~\ref{sec:CombiningSeeds}) consider these edge cases.

\paragraph{\bf Symmetry in preferences}
For the purpose of mutual acceptability, sometimes the user may require that, for every two agents $a_1$ and $a_2$, $a_1$ is the preference list of $a_2$ iff $a_2$ is the preference list of $a_1$. To ensure this, we add the following constraint to our seed generating ASP program.
\[
\lar  \ii{arank}(a_1,a_2,\_), not \;\; \ii{arank}(a_2,a_1,\_). \quad (1{\leq} a,a_1,a_2 {\leq} n;\ a_1 {\neq} a_2)
\]

Our seed \& combine algorithm (Section~\ref{sec:CombiningSeeds}) consider this possibility of generating symmetric preferences upon the user's request.

%%%%%%%%%%%%%%%%%%%%%%%%%%%%%%%%%%%%%%%%%%%%%%%%%%%%%%%%%%%%%%%%%%%%%%%%%%%
\section{Experimental Evaluations}

\paragraph{\bf Objectives} Recall that our study has been mainly motivated by Prosser's observation~(\citeyear{Prosser14}), described in Challenge~1: the SRI instances generated by the random instance generator of Prosser's software~\citep{sricp2019}, based on Mertens' idea~(\citeyear{Mertens2005}), have small numbers of stable matchings, and thus they do not sufficiently illustrate the exponential asymptotic complexity of an enumeration-based method (i.e., which first computes all stable matchings and then finds an egalitarian solution from among them) to solve Egalitarian SRI.  

Therefore, we have first conducted experiments to better understand the applicability and usefulness of our random instance generator from this perspective, and to answer the following questions:

\begin{itemize}
\item[(Q1)] Does our method generate SRI instances with many solutions, so as to illustrate the NP-hardness of Egalitarian SRI empirically with an enumeration-based method? 
\item[(Q2)] 
How does our random instance generator compare to the random instance generator of Prosser's software, in terms of the number of stable matchings of the instances they generate?
\end{itemize}

Next, we have conducted experiments to better understand the applicability and usefulness of our instance generator for empirical evaluation of an ASP-based method: 

\begin{itemize}
\item[(Q3)] Are the SRI instances generated by our method difficult for the ASP-based method of \citeauthor{Fidan20}~(\citeyear{Fidan20}) to solve Egalitarian SRI?
\end{itemize}

\paragraph{\bf Setup}
We have implemented our algorithm in Python (version 3.12.0), utilizing \clingo\ (version 5.4.0).

We have used Prosser's software~\citep{sricp2019} to generate random instances based on Erdös-Renyi model~(\citeyear{Erdos60}). We have also used Prosser's software to compute all stable matchings of SRI instances. This software utilizes the constraint programming system~\choco\ (version 2.1.5) via a Java wrapper (JDK 11.0.24). 

We have conducted our experiments on a laptop computer that has a 64-bit Ubuntu 20.04 as an operating system, and a 2.40 GHz x64-based processor with 8 GB RAM. 

\paragraph{\bf Benchmarks} 
We have used two sorts of SRI instances. 
\begin{itemize}
\item Existing benchmarks: 
We have considered the SRI benchmark instances that were earlier generated by \citeauthor{Fidan20}~(\citeyear{Fidan20}) using Prosser's software~\citep{sricp2019}, for an empirical analysis of various methods to solve SRI. This set contains, for each number $n{=}20, 40, 60, 80, 100$ of agents and for each completeness degree of $25\%, 50\%, 75\%, 100\%$, 20 SRI instances.\footnote{The completeness degree is defined~\citep{Fidan20} as $p\time 100$, where $p$ denotes the probability of the mutual acceptability of pairs of agents in a random graph ensemble $G(n,p)$ according to the Erdös-Renyi model~(\citeyear{Erdos60}). Note that the incompleteness probability $p_1$ used in our seed \& combine method equals to $1-p.$}

\item New benchmarks:
For every number $n{=}20, 40, 60, 80, 100$ of agents and for every incompleteness degree~$p_1{=}0,0.25,0.5,0.75$, we have generated new 20 SRI benchmark instances, using our seed \& combine method. For each SRI instance; 1) for every 20 agents, 3 different satisfiable seed instances are generated with $n=8,8,6$ and $k=6,6,2$, respectively, and $m=n-1$, and 2) these satisfiable seed instances are combined over every pair of agents, with $p_1$ and $p_2=0$.
\end{itemize}

\paragraph{\bf Experiments 1} 
In our first set of experiments, our goal to provide answers to questions Q1 and Q2. For these experiments, we have considered the first phase of an enumeration-based method only: computation of all stable matchings.\footnote{Recall that the second-phase an enumeration-based method finds an egalitarian stable matching from among all these solutions. The second phase is not included in the evaluations of the first set of experiments, as the first phase is sufficient to answer questions Q1 and Q2.} 

In this set of experiments, since Prosser considered the SR problem in his study~(\citeyear{Prosser14}), we have used only the SR instances from the existing (resp. the new) benchmark set, i.e., with completeness degree of 100\% (resp. with incompleteness probability $p_1{=}0$). 

For each number $n$ of agents; for each of the 20 SR instances with $n$ agents, we have used Prosser's software (with \choco) to find all stable matchings with a time threshold of 200  seconds in total.   For each satisfiable SR instance that could be solved within the time threshold, Prosser's software returns all the stable matchings it computes, and the total number of search nodes it examines. For each such satisfiable and solved SR instance, the CPU time (in seconds) to compute all stable matchings is also noted.

Table \ref{tab:FindingAllSolutions} illustrates the results of our experiments. For each number $n$ of agents, the number of satisfiable SR instances (out of 20 SR instances) that are solved within the time threshold is reported (\#SI). 

The average values are reported for the SR instances that are satisfiable and solved within the time threshold: the average number of stable matchings (\#SM), the average CPU time in seconds to compute all stable matchings (Time), and the average number of search nodes examined (Nodes).

In our experiments with the existing benchmarks, for each $n$, out of 20 SR instances, only some of them have stable matchings, and all satisfiable instances could be solved within the time threshold. 
For instance, for $n{=}100$ agents, 10 SR instances (out of 20 SR instances) have stable matchings; for each instance, all its stable matchings are computed within the time threshold. These 10 instances have in average 3.2 stable matchings. It takes 0.02 seconds in average to compute all stable matchings for these 10 instances. The average number of nodes is 7.1, and it denotes the average search effort.

From this table, one can see that, indeed, as observed by Prosser~(\citeyear{Prosser14}), the existing benchmarks do not show the hardness of Egalitarian SR: the average number of stable matchings is small, and thus it takes less than 1 second to find them all. In fact, the computation time almost never changes when the instance size increases. 

On the other hand, in the new benchmark, all SR instances are satisfiable, and  all satisfiable instances (except for the ones with $n{=}100$) are solved within the time threshold. 

In Table \ref{tab:FindingAllSolutions}, we observe that the average number $\#SM$ of stable matchings increases significantly, as the instance size $n$ increases.  For instance, when the number $n$ of agents increases from 60 to 80 (by 1.33 times), the average number of stable matchings increases from 373248 to 27097804 (by 72 times). In addition, the average CPU time increases from 2.3 seconds to 149.6 seconds (by 65 times), and the search effort (indicated by the number of examined nodes in search) increases by 72 times. 

As noted above, in the new benchmark, none of the 20 SR instances with $n{=}100$ could be solved within the time threshold. A solution to one of the instances could be found in more than 3 hours. For this instance, we have computed an estimate for \#SM as $65{\times}10^9$, using the approximate model counter {\sc ApproxMC}~\citep{ChakrabortyMV13}\footnote{{\sc ApproxMC}: \url{https://github.com/meelgroup/approxmc}} over the corresponding SAT formulation of SRTI.

Therefore, these experiments provide a positive answer to our question Q1: 
{\it
    Our method generates SRI instances with too many solutions, and can illustrate the NP-hardness of Egalitarian SRI empirically.
}

For question Q2, our experiments provide the following answer: 
{\it
Our random instance generator produces better instances, in terms of their number of stable matchings, compared to the random instance generator of Prosser's software.
}

\begin{table}[h!]
\centering
\caption{Results of computing all stable matchings, using Prosser's software with \choco. For $n$ agents, out of 20 SR instances; the number \#SI of instances with a solution, the average number \#SM of stable matchings, the average CPU time (in seconds) to compute all these stable matchings for the satisfiable instances, and the average number of search nodes are reported.}
\label{tab:FindingAllSolutions}
{\tablefont\begin{tabular}{@{\extracolsep{\fill}}lrrrrrrrr}
\topline
 & \multicolumn{4}{c}{Existing Benchmarks} & \multicolumn{4}{c}{New Benchmarks} \\
$n$ & Time & Nodes & \#SI & \#SM &  Time & Nodes & \#SI & \#SM
\midline
20  & 0.01  & 2.73  & 15 & 1.4 & 0.02  & 130 & 20 & 72 \\
40  & 0.02  & 4.53  & 15 & 2.33 & 0.12  & 9418 & 20 & 5184 \\
60  & 0.02  & 4.35  & 14 & 1.71 & 2.3   & 678154 & 20 & 373248 \\
80  & 0.02  & 5.69  & 13 & 2.6 & 149.63 & 49051094 & 20 & 27097804 \\
100 & 0.02  & 7.1  & 10 & 3.2 & TO  & -- & 0 & -- \\
\botline
\end{tabular}}
\footnotesize{TO: timeout, i.e., no solution was found for any of the 20 instances in 200 seconds.}
\end{table}

\paragraph{\bf Experiments 2} 
In our second set of experiments, the goal is to provide an answer to question Q3. For these experiments, we have considered computation of an egalitarian stable matching without enumeration.  We have used the ASP formulation of \citeauthor{Fidan20}~(\citeyear{Fidan20}) to solve Egalitarian SRI, with the existing and new benchmark instances, using \clingo. 

Table \ref{tab:ModelCounts} illustrates, for each instance size $n{=}20,40,60,80,100$ and for every incompleteness probability $p_1{=}0,0.25,0.5,0.75$ (i.e., for every mutual acceptability probability $1-p_1$), the number of SRI instances that have a solution, and the average CPU time (in seconds) to compute an egalitarian stable matchings for these instances.

We can observe from these results that the instances generated by our seed \& combine method require more computation time for larger instances. For instance, for $n{=}80,100$, these instances cannot even be solved with the ASP-based method. 

Therefore, these experiments provide a positive answer to our question Q3: 
{\it
    The SRI instances generated by our seed \& combine method are also difficult for the ASP-based method of \citeauthor{Fidan20}~(\citeyear{Fidan20}) to solve Egalitarian SRI.
}

\begin{table}[h!]
\centering
\begin{threeparttable}
\caption{Results of computing an egalitarian stable matching, using the ASP-based method of Fidan et al. with \clingo. For $n$ agents, incompleteness probability $p_1$, out of 20 SR instances; the number \#SI of instances with a stable matching, and the average CPU time (in seconds) to compute an egalitarian solution for these satisfiable instances are reported. }
\label{tab:ModelCounts}
%\resizebox{\textwidth}{!}
{\tablefont\begin{tabular}{@{\extracolsep{\fill}}llrrrrrr}
%\begin{tabular}{llrrrr}
\topline
    &     & \multicolumn{2}{c}{Existing Benchmarks} & \multicolumn{2}{c}{New Benchmarks} \\
$n$ & $p_1$ &              \#SI & Time                &            \#SI & Time             \\
\midline
\multirow{4}{*}[3pt]{20}& 0.00  & 15   & 0.06 & 20  & 0.09\\
            & 0.25  & 18   & 0.02 & 20  & 0.03\\
            & 0.50  & 17   & 0.01 & 20  & 0.01\\
            & 0.75  & 20   & 0.00 & 20  & 0.00\\
\hline
\multirow{4}{*}[3pt]{40}& 0.00  & 15   & 1.24 & 20  & 1.98\\
            & 0.25  & 14   & 0.51 & 20  & 0.45\\
            & 0.50  & 12   & 0.15 & 20  & 0.16\\
            & 0.75  & 11   & 0.02 & 20  & 0.06\\
\hline
\multirow{4}{*}[3pt]{60}& 0.00  & 14   & 7.59 & 20  & 16.71\\
            & 0.25  & 13   & 2.99 & 20  & 10.45\\
            & 0.50  & 16   & 0.86 & 20  & 9.95\\
            & 0.75  & 10   & 0.1 & 20  & 12.2\\
\hline
\multirow{4}{*}[3pt]{80}& 0.00  & 13   & 29.37 & 20  & {TO}\\
            & 0.25  & 8   & 10.76 & 20  & 85.72[1]{*}\\
            & 0.50  & 13   & 2.99 & 20  & {TO}\\
            & 0.75  & 13   & 0.37 & 20  & {TO}\\
\hline
\multirow{4}{*}[3pt]{100}& 0.00  & 10   & 93.18 & 20  & {TO}\\
           & 0.25  & 13   & 31.42 & 20  & {TO}\\
           & 0.50  & 12   & 7.82 & 20 & {TO}\\
           & 0.75  & 14   & 0.91 & 20  & {TO}\\
\botline
\end{tabular}}
\footnotesize{TO: timeout, i.e., no solution was found for any of the 20 instances in 200 seconds.}\\
\footnotesize{[1]*: only 1 instance out of 20 instances was solved in 200 seconds.}
\end{threeparttable}
\end{table}

\paragraph{\bf Repository} The implementation of our method is available at {\small\url{https://github.com/baturayyilmaz-su/GIG}}, and the benchmarks used in our experiments discussed above are available at {\small\url{https://github.com/baturayyilmaz-su/ICLP_Experiments/}}.

%%%%%%%%%%%%%%%%%%%%%%%%%%%%%%%%%%%%%%%%%%%%%%%%%%%%%%%%%%%%%%%%%%%%%%%%%%%
\section{Discussion}

We have introduced a two-step seed \& combine method to generate satisfiable benchmark instances for SRI possibly with Ties, that have large numbers of (e.g., over $10^6$) stable matchings, and thus it would not be practical to find an egalitarian stable matching by enumerating all stable matchings. 

We have utilized ASP to generate small satisfiable seed instances with multiple stable matchings, and introduced a novel algorithm to construct large instances from these seeds ensuring a lower bound on the number of stable matchings.  

Based on experimental evaluations, we have observed that our seed \& combine method indeed generates instances with significantly large number of stable matchings, compared with the instances generated by the existing graph-based approaches. We have also observed that, these instances are more difficult for the ASP-based method that is not based on enumeration of all stable matchings.

Since our method for generating satisfiable seed instances relies on an elaboration tolerant representation of SRI in ASP, and our method for populating and combining satisfiable seed instances preserves the stable matchings of these seed instances, our overall method can be easily used to generate satisfiable seed instances for other nonpartite or bipartite one-to-one stable matching problems with large numbers of stable matchings. 

For example, the stable marriage problem (SM) is a bipartite one-to-one stable matching problem. SM is a special case of SR: the set of agents consists of two types of agents (i.e., men and women), and no agent prefers another agent of the same type. So we can generate small satisfiable seed SMTI instances using our ASP formulation for generating SRTI instances, by adding relevant rules and constraints.   We can populate and combine satisfiable seed SMTI instances using our method for obtaining larger SRTI instances, by ensuring that an agent is not included in the preference of another agent of the same type. In fact, our implementation allows generation of SMTI instances with large number of stable matchings, with a command line option.

We believe that empirical evaluations of computational methods for matching problems is important to better understand the practical and theoretical challenges of these problems. Empirical evaluations also play an important role in the investigation of the applicability and usefulness of these computational methods in real-world applications. Our study of generating benchmark instances for one-to-one matching problems, that have large number of stable matchings, contributes to design and implementation of empirical evaluations of methods for matching problems, not only by providing models and methods but also by providing a tool for future studies.

\paragraph{Acknowledgments.} We thank Muge Fidan and Patrick Prosser for useful discussions and suggestions on benchmark instance generation for matching problems and empirical evaluations of matching methods. We thank the anonymous reviewers for their useful suggestions to improve our paper.

%%%%%%%%%%%%%%%%%%%%%%%%%%%%%%%%%%%%%%%%%%%%%%%%%%%%%%%%%%%%%%%%%%%%%%%%%%%

\FloatBarrier % force placement of all floats before here
\bibliographystyle{tlplike}
%\bibliography{baturayBib}

\end{document}